# RMAWS: An Hybrid Architecture for achieving Web Services Reliability in Mobile Cloud Computing

First A. Amr S.Abdelfattah, Second B. Tamer Abdelkader, and Third C. EI-Sayed M. EI-Horbaty

*Abstract*—The intermittent wireless connectivity in Mobiles limits the spread of mobile applications usage over the web, such that the web services are the independent protocol that used to achieve the mobile connectivity with the cloud services. Achieving the web service reliability results in two aspects. The first is low communication overhead and retrieving the appropriate response to prevent the duplicate request execution. The second is overcoming the request time out problem that is one of the most effected issues in the mobile experience. This paper proposes Reliable Mobile Agent and Web Socket approach (RMAWS) that achieves the reliable web services consumption. The enhanced architecture is hybrid between the mobile agent approach and web socket open connection communication protocol. This approach recovers the data lose because of the intermittent connections, overcomes the Time-out problem, and enhances the mobile experience by achieving this service consumption reliability.

*Keywords*—Web service; Reliable request; Mobile Cloud Computing; Mobile Agent Architecture; Web socket.

## I. INTRODUCTION

The recent explosion of the cloud computing is facilitating the deployment of web services, such that the web services act as self-contained components, which are published, located and invoked over the Web. The web services are the perfect way to provide a standard platform and operating system independent mechanism. This triggered the wide spread usage of mobile applications over the Web.

The web services deployment is facilitated with the appearance of cloud computing and its recent explosion. Cloud Computing improves scalability and consistency of services and data, and uses web services for connections, which facilitates the deployment of the mobile applications. [1]. Most of these cloud oriented services and data are deployed as web services that are network-oriented applications [2]. The synchronization between a mobile client and a web service is achieved through initiating a conversation in a request response pattern.

The consumption of web services through mobile client prefers the choice of Representational State Transfer (REST) service type. This is because REST architecture is fundamentally client-server architecture, and is designed to use a stateless communication protocol, typically HTTP. In REST architecture, clients and servers exchange representations of resources using a standardized interface and protocol. These characteristics encourage REST applications to be simple and lightweight. Therefore, regarding the scope of reliability, RESTFUL services overcome Simple Object Access Protocol (SOAP) services limitations and achieve better results, especially in mobile communications [3]-[5]. REST services use HTTP request and response, which means that a mobile device connected with the Internet can access the service without additional overhead, unlike SOAP web services [6]. Therefore, the consumption of these services and pool of data and information is affecting smart phones to gradually become the effective client platform to consume.

Mobile Cloud Computing (MCC) is the combination of cloud computing, mobile computing and wireless networking to bring rich computational resources to mobile users, network operators, as well as cloud computing providers [7].

The uncertainty of mobile connectivity results in less satisfaction for the mobile user. In addition the network bandwidth limitation and unreliable wireless communication are decreasing the overall support for web service consumption on mobile devices [8]. The limitations are listed in the following points:

- Client Connection loss: The mobile clients have intermittent connection because of their mobility. They can be momentarily removed from the connected network and later join the available network [6].
- Service Connection Loss: The Cloud/Server may lose the connection and their deployed web services become unreachable from clients.
- Server Error: The Cloud services may temporarily suffer from unexpected error, which may be produced because of high load requests, or system environment issues. This will overload the mobile client to reconstruct and resend the request later.
- Bandwidth limitations: Cellular networks have a very limited bandwidth that may cause slow service consumption or request timeout response.
- Reliability: The public Internet is unreliable, such that if a client calls a web service and doesn't get a response within the timeout period, the client doesn't know whether the web service request was successfully received, was lost in the Internet before reaching the server, or was partially processed. In the case the application retries the operation and resends the request, it may be duplicated or cause an error, such as two orders entered or two credit card charges.
- Longer Transaction time: Web service consumption will be slower because of the HTTP overhead, the XML overhead, and the network overhead to a remote server. Therefore the differences in performance need to be factored into the application architecture to prevent unexpectedly poor performance due to the latency of the web services consumption failures.

The service time out problem is one of the most effected issues in the mobile experience, such that the services response size, database relations' communication time, and the required computations in the services are continuously increasing corresponding to the application usage during the time, which causes repeat timeout through the consumption of these services.

In a highly distributed system where web services are scattered across multiple platforms, three system guarantees are required: Consistency of the data, Availability of the system/data, and Partition tolerance to fault. However, the "CAP theorem" [9] states that at most only two of the three can be guaranteed simultaneously. In distributed mobile systems where the mobile node is employed as the client platform of the web services, partition tolerance is a given because of the intermittent connectivity losses. This means we are forced to choose between Availability and Consistency [1].

There are two approaches achieve the reliable web service consumption: Middleware approach and Mobile Agent (MA) approach. Both of them focus on representing the user in wireless networks. They perform the requests using the user arguments and return the results to the user in his/her next online time.

The web socket Protocol is an independent TCP-based protocol. It makes more interaction between a client and a server, facilitates the real-time data transfer from and to the server. This is made possible by providing a standardized way for the server to send content to the client without being solicited by the client, and allows messages to be passed back and forth while keeping the connection open [10].

The proposed approach integrates the mobile agent approach and web socket open connection protocol, in order to ensure the request execution, prevent the duplicate request execution, and overcome the request timeout problem to the cloud services. The mobile agent uses the web socket open connection to send the response in case of request time-out occurs. This time-out problem tightly coupled with the increasing of the service response size, database relations' communication time, and the required computations in the service, which occurs because of the application usage increasing.

The rest of this paper is organized as follows: Section II contains an overview of the related work. The proposed RMAWS architecture and its components detail is explained in section III. The RMAWS protocol analysis is figured in section IV. The environment setup is shown in section V. Section VI discusses and analyzes the conducted results. The conclusion and future work are given in section VII.

## II. RELATED WORK

The mobile agent approach is applied in more than context with different purpose. The mobile agent is tailored to represent the user presentence for achieving the required task. The authors focused on the main point of caching technique to enable the user recovers the response in the second connects to the service.

In [11], the author used the REST requests as the web service type that deployed over cloud. The mobile agent is constructed once the mobile client connects to the cloud service. The agent executes the request with the behave of the user, then the agent checks the user availability. The mobile agent either sends the response result to the mobile client if it was available or caches the results till the mobile reconnects to the service next time.

The following related Mobile Agent approach is used to represent the user in wireless networks. They perform the requests using the user arguments even in case of the user absence and return the results to the user if s/he is online or the next connecting time, the user receives the results, as shown in Fig. 1.

The proposed approach is considering the mobile agent architecture hybrid with the web socket protocol to enhance the communication cycle by: identifying the request itself rather than the mobile client in order to enable the multi-request per client and multiple caches per client, and overcoming the time out request based on the proposed technique. This approach enhances the behavior of the request cycle rather than modifies the cloud service original structure, to achieve the REST web services reliability through mobile cloud computing.

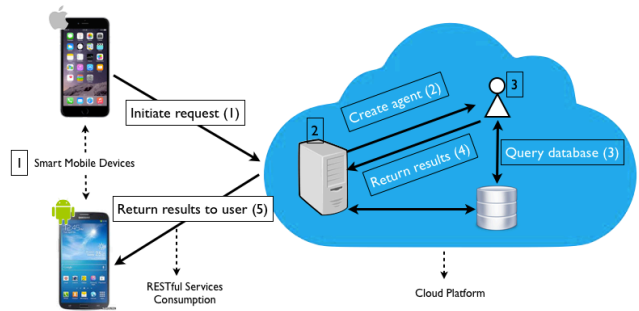

Fig. 1 Mobile agent Architecture and the communication process between its components

### A. MA Architecture

Mobile agents architecture as shown in Fig. 3, consists of the following components:

1. A Mobile device that has Internet access.
2. A Web application that contains web services connected to a database to store the needed data.
3. A Mobile Agent that represents the user created by the system that consumes the services repository to get the response.

### B. MA Advantages

Mobile agents advantages can be summarized as following:

- Overcome the disconnection problem, it enriches the RESTFUL services using the mobile agent that represents the user in wireless networks.
- Doesn't need the user persistence in the entire session, so if the user disconnected due to network connection failure or empty battery, the user can get the results in the next time s/he connect to the system again.

### C. MA Limitations

Mobile agents limitations can be discussed as following:

- Agent gets and invokes only one service from the service repository.
- Multi-agents are needed in parallel to measure using parallelism on the response time if the user requested more than a service at the same time.
- The security layer is needed to be added to the request and the response to be used safely in confidential applications.

### III. PROPOSED RELIABLE APPROACH USING MOBILE AGENT AND WEB SOCKET (RMAWS)

The proposed approach integrates the mobile agent approach with its architecture with another protocol for communication, which is the web socket protocol. This integration allocates the mobile agent to consume the required web service with behave of the mobile client with its arguments, and offers the web sockets to represent the open connection protocol between the cloud server and the mobile client.

The proposed approach is based on the architecture that shown in Fig. 2, contains four main components:

1. **Mobile Cloud Consumer** is a mobile client component responsible for:
   - Constructing the request with its appropriate attributes to send it to the Cloud Service Component layer.
   - Handling the request communication cycle between the mobile layer and the service layer.
   - Managing the client side web socket communication.
   - Receiving the response and notifying the client with the appropriate result based on the service response and the web socket data communication.
2. **Cloud Service Component** is the component that responsible for:
   - Receiving the request from the Mobile Cloud Consumer, and constructing the appropriate response regarding the sent request attributes.

- Create and initiate the mobile agent to perform the appropriate task regarding to the sent request.
- Managing the server side web socket communication with the Cloud Client Consumer.
- Handling the response to be sent to the Mobile Cloud Consumer.
3. **Web socket protocol** is a standard communication way for the server to send content to the client without being solicited by the client. It plays the most important role to overcome the request time-out problem, such that it allows messages to be passed back and forth while keeping the connection open during the service communication time, and it is responsible for:
- Achieving the other way of communication between the Mobile Cloud Consumer and the Cloud Service Component, beside the required REST service communication.
- Sending the actual response to the Mobile Cloud Consumer once this response returned from the cloud service in case of time-out problem is occurred.
4. **Mobile Agent** is the component that represents the user presentence to perform the required task even in case of the user absence, it responsible for:
- Querying the persistence storage to perform the required task.
- Caching the result in the persistence storage in case of the user absence while the results are fetched.
- Retrieving the required cached data to be sent to the Mobile Cloud consumer in its next communication.

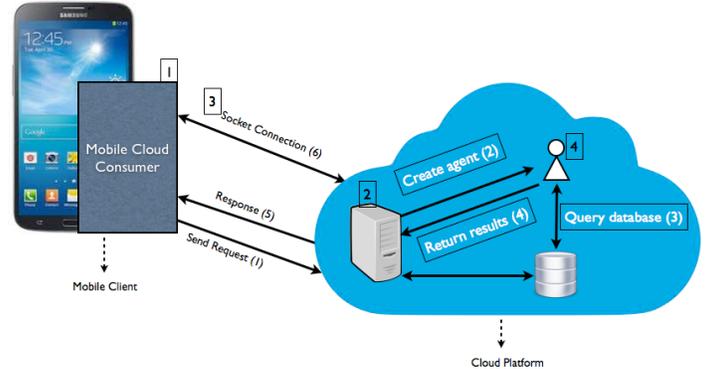

Fig. 2 RMAWS Architecture

IV. RMAWS PROTOCOL

The RMAWS architecture achieves the reliability for the heavy cloud service while overcoming the time-out problem, and preventing the duplicate executing for a specific request.

The RMAWS sequence diagram is shown in Fig. 3 shows the communication cycle that is proposed to achieve the service consumption reliability using Mobile Agent and Web socket.

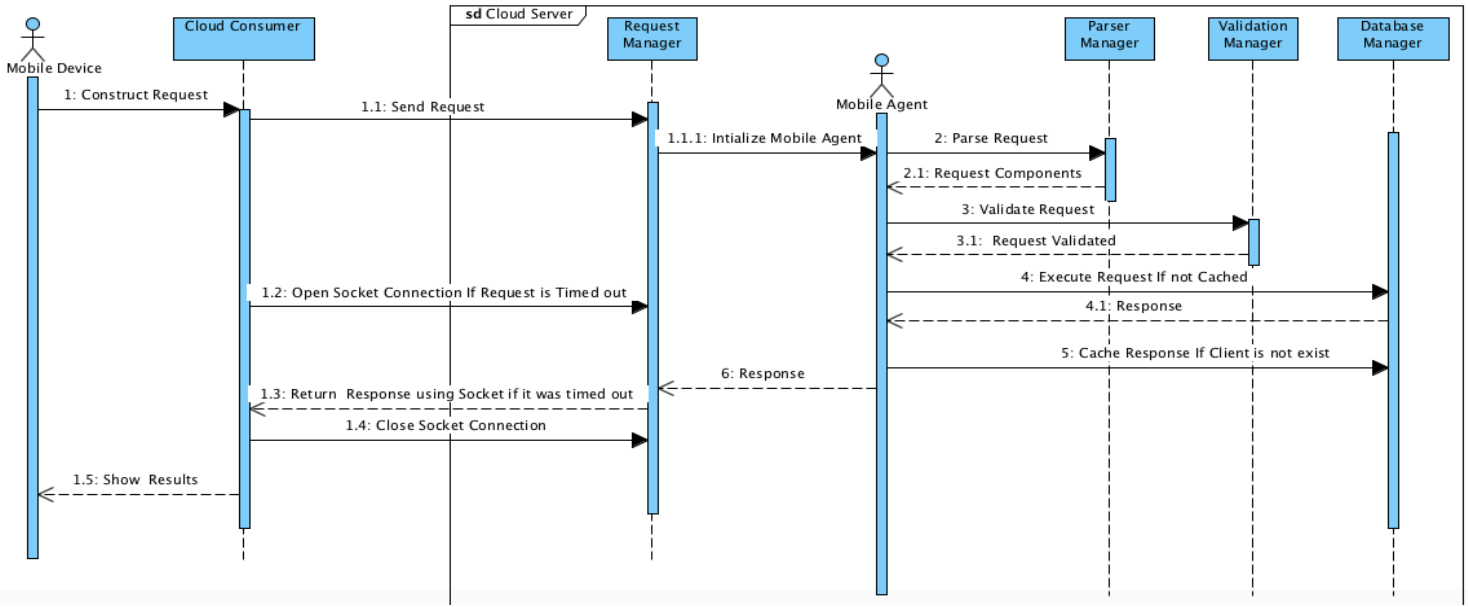

Fig. 3 RMAWS Protocol Sequence

The RMAWS protocol applies the following algorithm,
- [Mobile Device] The user selects a specific request to execute.
- [Cloud Consumer] Construct the REST request and add its additional attributes (1), such as: request_id, and is_forced.
- [Cloud Consumer] Send the request to the specific Cloud Service.
- [Request Manager] Create and initialize the Mobile Agent to perform the request.
- [Parser Manager] Parse the request to extract the embedded attributes.
- [Validation Manager] Validate the request to ensure the correctness and authorization attributes.
- [Database Manager]
    - Check if this request is cached before to return its response.
    - "If not cached" Execute the request and return the response.
- [Cloud Consumer] "If request is timed-out" Open socket connection with the cloud server.
- [Mobile Agent] Returns the response to the Request Manager.
- [Mobile Agent]
    - Check if the user is still exists in the system to receive the response.
    - "If Client isn't exist" Send the response using Web socket communication with the client.
- [Database Manager] "If client is not exist" Cache the response attached with its request identifier.
- [Cloud Consumer] "If socket connect is opened" Close the web socket connection.
- [Mobile Device] Notify the user with the returned results.

The above algorithm contains the following:
(1) The request additional attributes: They are attributes that control some important behaviors in this process, such as:
   1. request_id: is the unique identifier over all requests in the system, to distinguish the responses returned through the web socket connection from multiple requests. It consists of two main parts Identifies the request:
      a. **Basic Part** is the auto generated unique key that consists of subparts that ensure its distinct property, such as mobile device manufacture number, the request timestamp, and the requested service name.
      b. **Additional Attributes Part** is the other part that includes labeled attributes that change the Mobile Agent behavior regarding the received request such as the request number of trials, and forced attribute.
   2. is_forced: The forced attributes that force the Mobile Agent to use the succeeded cached results, or re-execute the request each time.

## V. ENVIRONMENT SETUP

The mentioned RMAWS architecture is applied in the following environment in Table I. It shows the programming languages, the tools that used in building this enhanced architecture, and the cloud service provider that is used to deploy the services.

| **Cloud Service Component** | | | |
| --- | --- | --- | --- |
| **Language & Frameworks** | **Web service & Web socket** | **Mobile Agent** | **Database &Text Format** |
| Java Enterprice Edition & Hibernate Framework | RESTFull (JAX-RS) & Java$^{TM}$ API for WebSocket (JSR 356) | Java Agent Development Framework (JADE) [12] | MySQL & JSON Format |
| **The mobile application and the Mobile Cloud Consumer** | | | |
| **Platform** | **Web socket** | | **Text Format** |
| Android | Java WebSockets [13] | | JSON |
| **Service Deployment** | | | |
| **Cloud Service Provider** | | **Application Server** | |
| Openshift Cloud Service Provider | | Glassfish Application server | |

Table I Environment Setup

## VI. RESULTS AND ANALYSIS

The measurements are conducted as a comparison between the direct connection to the cloud service versus the connection through the proposed RMAWS components. The proposed architecture handles affected cases corresponding reliability concept as mentioned in the above sections. Moreover the performance comparison will be considered to discuss the affected factors related to the RMAWS approach. The following factors are considered as the most affective factors regarding to the mobile computing scale:

- Request size: The size of the request body that constructs and send in the mobile client, it will be affected because of the additional attributes attached with the request.
- Response size: The size of the response body that it is received in the mobile client, it won't vary because there are no additional properties need to be added from the mobile agent for any reason.
- Consuming time: The time taken from sending the request till receiving its response may be longer because of the mobile agent initiation process cost time.

| Direct Cloud Connection | | | RMAWS Connection | | |
|---|---|---|---|---|---|
| **Request Size (Byte)** | **Response Size (Byte)** | **Consuming Time (Second)** | **Request Size (Byte)** | **Response Size (Byte)** | **Consuming Time (Second)** |
| 25 | 191745 (~ 191KB) | 2 | 251 | 191745 (~191KB) | 2 |
|  | 2167000 (~2MB) | 21 – 23 |  | 2167000 (~2MB) | 22- 25 |
|  | 4592949 (~4.5MB) | 44 |  | 4592949 (~4.5MB) | 46 |
|  | 6828571 (~7MB) | 65 – 66 |  | 6828571 (~7MB) | 65 – 68 |
| 55 | 5 | > 0 | 281 | 5 | > 0 |

Table II Performance Measurements

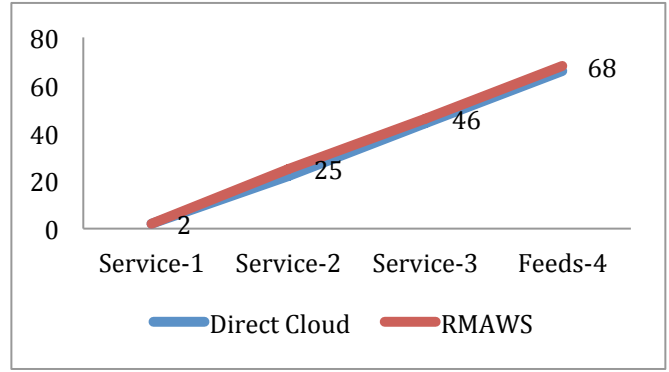

Fig. 6 Consuming Time (Seconds) relation between direct cloud and RMAWS connection

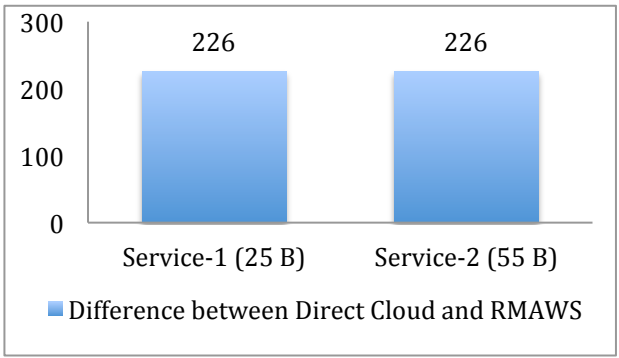

Fig. 4 Request Size (Byte) relation between direct cloud and RMAWS connection

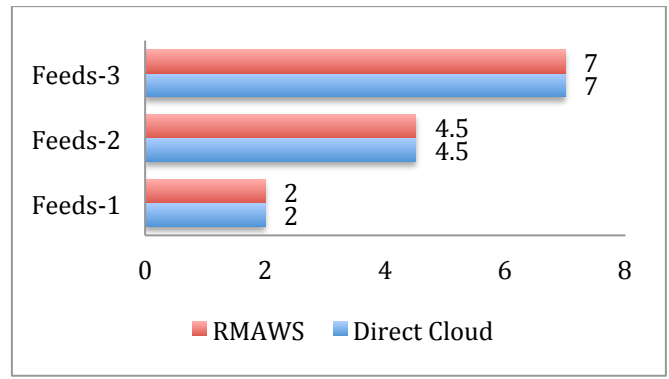

Fig. 5 Response Size (MB) relation between direct cloud and RMAWS connection

Table II, figures 4, 5, 6, show the performance measurements through RMAWS components and through direct cloud service connection.

The measurements indicate regarding the request size factor, the RMAWS adds additional 226 bytes to the request. These bytes used for the additional attributes discussed in Section IV. Although this number of additional bytes, this cost is low and its complexity is $O(1)$ such that it is increasing rate is constant regardless the original request size.

The response size factor is not changed, the mobile agent returns the response as it is fetched. Therefore, we avoid using transformation overhead in the response structure.

The consuming time is one of the most important factors for the mobile applications, which require quick responses. Regarding the different response sizes, the measurements show that time may be the same in both environments or may be slightly vary about 1 to 3 seconds. This variation is because the initialization process for the Mobile Agent (that can be differ corresponding to the mobile agent generator API), and its corresponding validation and checks.

In the other side, the proposed approach overcomes the Time-out problem that threats the cloud service reliability. It avoids the next retry technique to retrieve the cached data when the mobile client is received timed-out response. The proposed approach of Web socket open connection reduces the Mobile Cloud Consumer overhead that is because of the request re-construction and the network resource allocation to resend the REST request to get the cached response if it is returned and saved successfully in the persistence storage. From the cost perspective, it requires extra memory for this web socket communication, which is opened when the time-out problem is occurred.

## VII. CONCLUSION AND FUTURE WORK

The proposed Reliable Approach using Mobile Agent and Web Socket (RMAWS) achieves the reliability by integrating the reliable mobile agent approach with the Web socket open connection protocol. It is focuses on ensuring the request execution and preventing the duplicate request execution as a result to the intermittent mobile connection. In addition, it considers the most important factors for the mobile client such as the request size and the response size for mobile client data transmission limitations, and the service consuming time, which is critical for the mobile applications and their usability.

The cloud services consumption by the RMAWS architecture and its protocol guarantees the reliable service communication between the mobile client and the cloud service, while it avoids adding significant communication overhead compared with the traditional direct cloud service consumption with regard to the mentioned three factors.

The proposed architecture conducts a solution for the Time-out problem. This problem that is frequently occurred in real time applications, because the network decreasing transfer rate, and another factors that are continuously increasing corresponding to the application usage, such as: the response size, database relations' communication time, and the required computations in the services.

Future work will be invested in implementing an enhanced approach that reduces the network usage in the heavy computation services, and consider the middleware as a reliable approach to compare its behavior with this proposed approach. This future approach will mainly concern with reducing the consumption time overhead from the mobile side, and keeping the heavy services smoothly consumed without time-out failure.